%% file: main.tex
\documentclass[sigconf]{cidr-2025}

\usepackage{xcolor}
\usepackage{enumitem}
\usepackage{amsmath}
\usepackage{dblfloatfix}
\usepackage{cleveref}

\usepackage{subfiles}
\usepackage{blindtext} 
\usepackage{listings}
\usepackage{array}
\usepackage{soul}
\usepackage{xspace}
\usepackage[ruled,vlined,linesnumbered]{algorithm2e}

\usepackage{colortbl}
\usepackage{booktabs}
\usepackage{tikz}
\usetikzlibrary{positioning, shapes.multipart, calc, arrows.meta}
\usepackage{subcaption}
\usepackage{graphicx}
\usepackage{balance}
\usepackage{tabularx}
\graphicspath{{images/}{../images/}}
\usepackage{adjustbox}

\usepackage{pgf}
\usepackage{tikz}
\usetikzlibrary{patterns}
\usetikzlibrary{decorations.pathmorphing} 

\input{commands.tex}

\title{Flow with FlorDB: Incremental Context Maintenance for the Machine Learning Lifecycle}

\author{Rolando Garcia}
\affiliation{%
      \institution{UC Berkeley}
      \country{United States of America}
}
\email{rogarcia@berkeley.edu}

\author{Pragya Kallanagoudar}
\affiliation{%
      \institution{UC Berkeley}
      \country{United States of America}
}
\email{pkallanagoudar@berkeley.edu}

\author{Chithra Anand}
\affiliation{%
      \institution{UC Berkeley}
      \country{United States of America}
}
\email{chithra.rajan@berkeley.edu}

\author{Sarah E. Chasins}
\affiliation{%
      \institution{UC Berkeley}
      \country{United States of America}
}
\email{schasins@berkeley.edu}

\author{Joseph M. Hellerstein}
\affiliation{%
      \institution{UC Berkeley}
      \country{United States of America}
}
\email{hellerstein@berkeley.edu}

\author{Erin Michelle Turner Kerrison}
\affiliation{%
      \institution{UC Berkeley}
      \country{United States of America}
}
\email{kerrison@berkeley.edu}

\author{Aditya G. Parameswaran}
\affiliation{%
      \institution{UC Berkeley} 
      \country{United States of America}
}
\email{adityagp@berkeley.edu}

\begin{document}

\begin{abstract}
In this paper we present techniques to incrementally harvest and query arbitrary metadata from machine learning pipelines, without disrupting agile practices.
We center our approach on the developer-favored technique for generating metadata --- log statements --- leveraging the fact that logging creates context. We show how hindsight logging~\cite{garcia2020hindsight} allows such statements to be added and executed post-hoc, without requiring developer foresight. Relational views of incomplete metadata can be queried to dynamically materialize new metadata in bulk and on demand across multiple versions of workflows. This is done in a ``metadata later'' style, off the critical path of agile development. We realize these ideas in a system called FlorDB and demonstrate how the data context framework covers a range of both ad-hoc metadata as well as special cases treated today by bespoke feature stores and model repositories. Through a usage scenario---including both ML and human feedback---we illustrate how the component techniques come together to resolve classic software engineering trade-offs between agility and discipline.
\end{abstract}

\maketitle

\section{Introduction}
In the rapidly evolving field of Artificial Intelligence (AI) and Machine Learning (ML), the management of metadata has emerged as an enduring challenge~\cite{sculley2014machine, garcia2018context}. 
As machine learning models and their applications become increasingly integral to business and technology, the need for accurate and comprehensive metadata management has never been more critical. However, this requirement presents numerous difficulties, primarily due to the highly diverse mix of systems and artifacts involved in MLOps. Compounding this complexity is the necessity for robust feedback loops that span organizational boundaries—such as those between design and deployment teams—and extend over time.

\subsection{A Crisis of Metadata Management}
One persistent challenge is balancing agility with the rigor of upfront documentation or ``metadata first'' approaches. Detailed documentation and metadata are essential for reproducibility, debugging, and collaboration. Unfortunately, insisting on comprehensive metadata from the start can hinder the agility and speed that are often crucial in early development stages. 

The tradeoff between discipline and agility is an enduring open problem, evident in various contexts including the dichotomy between schema-first and NoSQL databases, and the contrast between data warehouses and data lakes. Schema-first and warehouse approaches offer discipline and structure but lack the flexibility needed for rapidly evolving requirements. Conversely, NoSQL and data lake approaches provide the necessary agility but can lead to inconsistency and disarray in metadata management.

To address this, two core goals can be established for resolving what has been considered an inherent conflict:

\begin{enumerate}
\item{\textbf{Goal 1. Agile Development Loop:}} 
Metadata capture should be gradually incorporated into the agile workflows of data scientists and MLEs without interference. For data scientists, metadata capture should fit naturally into their open-source development environment.
For MLEs, metadata should fit into standard tools for workflow management and scalability, such as relational databases and CI/CD pipelines, without requiring lock-in on yet another service for metadata management. 
    
\item{\textbf{Goal 2. Metadata on Demand:}} Metadata generation, like other features, should be able to evolve incrementally, improving organically with project needs. Ideally we want a ``metadata later'' approach that easily back-propagates metadata discipline into earlier versions of the project --- at just the time when someone regrets not having the metadata they need.
\end{enumerate}

\subsection{Goals and Contributions} 
Our work is built on FlorDB\footnote{\url{https://github.com/ucbrise/flor}}, the scope of which we significantly expanded to encompass the entire ML lifecycle. FlorDB now captures complete pipelines and their regular execution, rather than just individual tasks. These extensions apply to mechanisms managed by the FlorDB API, initially designed for multiversion hindsight logging but now adapted to include incremental context maintenance over workflows. 
Despite these extensions, FlorDB maintains API stability and backward compatibility with multiversion hindsight logging.

Key goals and contributions include:
\begin{enumerate}
\lstset{style=custompython}
\item{\textbf{Incremental Context Maintenance for Agile DevOps:}} Our system allows developers to log and analyze metadata in a standard, open, low-friction manner. Metadata can be captured naturally through Python \lstinline|log| statements as part of the development workflow, without imposing significant overhead. Subsequently, these log statements can be read directly as tabular data using standard Python \lstinline|dataframe|s, queried via Pandas or SQL, without requiring data wrangling.
    
\item{\textbf{Ad-Hoc Metadata on Demand Enabled by Hindsight Logging:}} FlorDB enables agile metadata evolution through multiversion hindsight logging, which operates as a record-replay mechanism. While traditional logging systems require predefined schemas, our approach captures sufficient execution state during recording, and allows the extraction of \emph{arbitrary expression values} derivable from that state during replay. This means metadata collection isn't limited to just what was explicitly recorded; developers can compute arbitrary new properties and relationships from the execution record after the fact. Like materializing a view, developers can declare what metadata they want to extract, but unlike views, the source material extends beyond stored data to encompass both the complete execution record and the universe of derivable properties. This powerful mechanism frees developers from ``metadata first'' constraints while enabling unbounded possibilities for post-hoc metadata computation and analysis.

\item{\textbf{Unified, Open Metadata for Machine Learning:}} Last, the open, standard approach of FlorDB simplifies and improves the abstraction of metadata, incorporating features from various bespoke ML metadata systems such as feature stores, model registries, and labeling systems into a unified and robust framework. 
\end{enumerate}

We demonstrate these contributions through a document intelligence use-case, highlighting the importance of context in streamlining development cycles and improving operational efficiency. 

\section{Multiversion Hindsight Logging}\label{sec:mvhl}

This section provides background on Flor and FlorDB, highlighting their evolution and key features in managing the machine learning lifecycle.
Flor, originally designed as a record-replay system for model training~\cite{garcia2020hindsight}, offers two main features: i) low-overhead adaptive checkpointing, minimizing computational resources during model training, and ii) low-latency replay from checkpoints, leveraging memoization and parallelism speed ups.
Flor's record-replay mechanism introduced the notion of hindsight logging, allowing for post-hoc and on-demand querying of effectively unbounded \emph {context}. 

We adopt the term ``context'' from the Ground project~\cite{hellerstein2017ground}, as a signifier for an all-encompassing view of metadata that goes beyond traditional relational metadata. Here, ``context'' includes anything that could be emitted by a log statement in any running process, be it an ML training job, a data wrangling script, a live inference server, a log processor handling usage feedback, or even an orchestration framework that knits these other tools together.

FlorDB extends Flor's capabilities by integrating automatic version control, adding a relational data model for querying logs, and cross-version logging statement propagation for multiversion hindsight logging~\cite{garcia2023multiversion}. Its relational model, accessible via \lstinline[style=custompython]{flor.dataframe}, maps individual logging statements into columns in a pivoted view. This approach facilitates easy tracking of changes over time.
In this paper (Section~\ref{sec:api}), we further extend FlorDB to support dataflow pipelines and manage feedback loops. This extension provides a seamless framework for capturing arbitrary context introduced in defining and executing complex, evolving ML pipelines. 

To clarify the concept of multiversion hindsight logging, imagine a scenario where a developer has run several versions of a machine learning pipeline but later realizes that certain metadata or context was not captured during those runs. Traditionally, retrieving this missing information would require re-running each version of the pipeline with the new logging statements inserted---a process that is both time-consuming and resource-intensive. FlorDB's multiversion hindsight logging offers a powerful alternative that minimizes both developer effort and compute resources. Developers can add the desired logging statements to the latest version of their code, and FlorDB will (a) inject these statements into the correct locations in all prior versions of the code, and (b) retroactively execute these statements across all those versions via incremental replay, without the need for full re-execution. The former, (a), is made possible via techniques adapted from code diffing~\cite{falleri2014fine}; the latter, (b), is made possible through a combination of differential execution and parallelism, allowing FlorDB to efficiently replay only the necessary parts of the pipeline to extract the new metadata. This ``magic trick'' enables developers to incrementally build up context and metadata after the fact, supporting agile practices by eliminating the need for foresight in logging. It resolves the classic trade-off between starting fast and refining over time by providing the flexibility to log now, and get data from the past. For technical details on how this is achieved, we refer readers to our prior work~\cite{garcia2023multiversion}.

\subsection{FlorDB Extended API}
\label{sec:api}
FlorDB's API captures metadata about the executing file, eliminating the need to restate dataflow dependencies; cross-executable dependencies declared in a typical workflow orchestration tool (Airflow, MLFlow, Make, etc) suffice.
By profiling runtime metadata, including the executed file's name, FlorDB remains agnostic to the choice of workflow management system, functioning seamlessly  without requiring refactoring of orchestration scripts.

The Flor API, as presented in Garcia et al. (2023)~\cite{garcia2023multiversion}, includes:

\begin{itemize}
    \lstset{style=custompython}
    \item \lstinline|flor.log(name: str, value: T) -> T|: Logs a value with a specified name, constructing a record with \texttt{projid}, \texttt{tstamp}, \texttt{filename}, and nesting dimensions defined by \lstinline|flor.loop|.
    
    \item \lstinline|flor.arg(name: str, default: T) -> T|: Reads command-line values or uses defaults, retrieving historical values during replay.
    
    \item \lstinline|flor.loop(name:str, vals:Iterable[T]) -> Iterable[T]|: \\
    A Python generator maintaining global state between iterations, useful for addressing \lstinline|flor.log| records and coordinating checkpoints.
    
    \item \lstinline|flor.checkpointing(kwargs: Dict) -> ContextManager|: \\
    A Python "context manager" defining objects for adaptive checkpointing at \lstinline|flor.loop| iteration boundaries.
    
    \item \lstinline|flor.dataframe(*args) -> pd.DataFrame|: Produces a Pandas DataFrame of log information, with columns corresponding to each argument in \lstinline|*args| plus dimension columns like \texttt{projid}, \texttt{tstamp}, \texttt{filename}.
\end{itemize}

To support user interactions in long-running web applications, such as ``Save \& Close'' buttons, FlorDB is further extended with the following API call:

\begin{itemize}
    \lstset{style=custompython}
    \item \lstinline|flor.commit() -> None|: An application-level transaction commit marker supporting visibility control for long-running processes. It writes a log file, commits changes to git, and increments the \texttt{tstamp}. This method is automatically invoked (via \texttt{atexit}) at the end of a Python execution.
\end{itemize}


    

\begin{figure}
  \centering
  \resizebox{1.1\columnwidth}{!}{
    \begin{tikzpicture}[node distance=2cm and 2cm, blacktable/.style={rectangle split, rectangle split parts=2, draw, align=center},
                    bluetable/.style={blacktable, fill=brewergreen},
                    redtable/.style={blacktable, fill=gray!20}, font=\sffamily\small]

                 \node[blacktable] (loops) {loops \nodepart{two}
                        \begin{tabular}{ll}
                              \small \underline{projid}:   & \small text     \\[-0.2em]
                              \small \underline{tstamp}:   & \small datetime \\[-0.2em]
                              \small \underline{filename}: & \small text     \\[-0.2em]
                              \small \underline{ctx\_id}:            & \small integer     \\[-0.2em]
                              \small parent\_ctx\_id:      & \small integer     \\[-0.2em]
                              \small loop\_name:        & \small text     \\[-0.2em]
                              \small loop\_iteration:   & \small integer  \\[-0.2em]
                              \small iteration\_value:  & \small text
                        \end{tabular}};

                  \node[blacktable, left=of loops] (logs) {logs \nodepart{two}
                        \begin{tabular}{ll}
                              \small \underline{projid}:   & \small text     \\[-0.2em]
                              \small \underline{tstamp}:   & \small datetime \\[-0.2em]
                              \small \underline{filename}: & \small text     \\[-0.2em]
                              \small \underline{ctx\_id}: & \small integer     \\[-0.2em]
                              \small \underline{value\_name}:     & \small text     \\[-0.2em]
                              \small value:    & \small text     \\[-0.2em]
                              \small value\_type:     & \small integer
                        \end{tabular}
                  };
                        
                  \node[redtable, above=of logs] (ts2vid) {ts2vid \nodepart{two}
                        \begin{tabular}{ll}
                              \small \underline{projid}: & \small text \\[-0.2em]
                              \small \underline{ts\_start}: & \small datetime \\[-0.2em]
                              \small ts\_end:   & \small datetime \\[-0.2em]
                              \small vid:       & \small text     \\[-0.2em]
                              \small root\_target & \small text 
                        \end{tabular}};
                  \node[redtable, right=of ts2vid] (repo) {git \nodepart{two}
                        \begin{tabular}{ll}
                              \small \underline{vid}:        & \small text    \\[-0.2em]
                              \small \underline{filename}:   & \small text    \\[-0.2em]
                              \small parent\_vid: & \small text \\[-0.2em]
                              \small contents:   & \small text
                        \end{tabular}};
                  \node[redtable, below=of logs] (obj_store) {obj\_store \nodepart{two}
                        \begin{tabular}{ll}
                              \small \underline{projid}:   & \small text     \\[-0.2em]
                              \small \underline{tstamp}:   & \small datetime \\[-0.2em]
                              \small \underline{filename}: & \small text     \\[-0.2em]
                              \small \underline{ctx\_id}: & \small integer     \\[-0.2em]
                              \small \underline{value\_name}:     & \small text     \\[-0.2em]
                              \small contents:    & \small blob    
                        \end{tabular}};
                    \node[redtable, below=of loops] (build_deps) {build\_deps \nodepart{two}
                        \begin{tabular}{ll}
                            \small \underline{vid}: & \small text \\[-0.2em]
                            \small \underline{target}: & \small text \\[-0.2em]
                            \small deps: & \small text\texttt{[]} \\[-0.2em]
                            \small cmds: & \small text\texttt{[]} \\[-0.2em]
                            \small cached: & \small bool 
                        \end{tabular}
                    };

                  \draw[one to omany] 
                  ($(logs.east)+(0,0pt)$) --
                  ($(loops.west)+(0,0pt)$);
  
                \draw[one to zeroone] 
                ($(loops.north)+(-25pt,0)$) 
                to [out=160,in=20,looseness=3] 
                ($(loops.north)+(25pt,0)$);

                \draw[one to omany] 
                ($(build_deps.north)+(0,0)$) 
                to [out=90,in=0,looseness=3] 
                ($(build_deps.east)+(0,10pt)$);

                  \draw[one to zeroone] ($(logs.south)-(0,0pt)$) -- ($(obj_store.north)-(0,0pt)$);
                  \draw[one to many]
                  ($(ts2vid.south)+(0,0pt)$) --($(logs.north)-(0,0pt)$);
                  
                  \draw[one to many] 
                  ($(ts2vid.east)-(0,0pt)$) -- ($(repo.west)+(0,0pt)$);

                \draw[one to one]
                ($(ts2vid.east)-(0,25pt)$) to [out=0,in=180,looseness=1]  ($(build_deps.west)+(0,7pt)$);
                  
                  \draw[one to zeroone] 
                ($(repo.north)+(-20pt,0)$) 
                to [out=160,in=20,looseness=3] 
                ($(repo.north)+(20pt,0)$);

            \end{tikzpicture}}

  \caption{Extended FlorDB data model in Crow's Foot notation. Basic tables denoted in white; virtual tables in gray.}
  \label{fig:datamodel}
\end{figure}

\section{Incremental Context Maintenance}
\lstset{style=custompython}
The ML lifecycle is characterized by numerous fast-changing components, where it is easy to lose track of essential metadata --- what we term \textit{context}. Context represents a comprehensive framework that captures the nature, origins, evolution, and functional significance of data and digital artifacts within an organization. It is metadata broadly conceived, extending beyond traditional database metadata to encompass the full spectrum of information necessary for understanding and managing projects.

We base our conceptualization of context on the ``ABCs of Context'' framework introduced by Hellerstein et al. (2017)~\cite{hellerstein2017ground}, which extends traditional database metadata to encompass a broader spectrum of information critical in ML applications. The framework includes:
\begin{enumerate}
    \item \textbf{Application Context} (the ``A''): Captures how raw data is interpreted, including schemas, checkpoints, and parameters. FlorDB captures application context through log statements, allowing developers to record pertinent information during execution.
    \item \textbf{Behavioral Context} (the ``B''): Tracks how data is created and used over time, relating to lineage or provenance. This context is often contained in build files, capturing dependencies and directed acyclic graphs (DAGs) of tasks. FlorDB captures behavioral context via its relational data model, associating each logging statement with its originating filename. This mapping reveals dataflow pathways through the codebase, enabling analysis of data transformations and lineage within the pipeline.
    \item \textbf{Change Context} (the ``C''): Manages version histories of both data and code. FlorDB manages change context using Git version control, ensuring that all runs and modifications are tracked and retrievable.
\end{enumerate}
This framework provides a structured approach to understanding and managing the rich tapestry of information that underpins real-world ML applications.


\begin{figure*}[t]
    \centering
    \begin{minipage}[b]{0.18\linewidth} 
        \begin{lstlisting}[style=makefile]
# Makefile

prep:
    python prep.py
    
infer: prep
    python infer.py
    
run: infer
    flask run
    
train: prep
    python train.py
        \end{lstlisting}
    \end{minipage}%
    \hspace*{-0.3cm}
    \begin{minipage}[b]{0.4\linewidth}
        \centering
        \includegraphics[height=4cm]{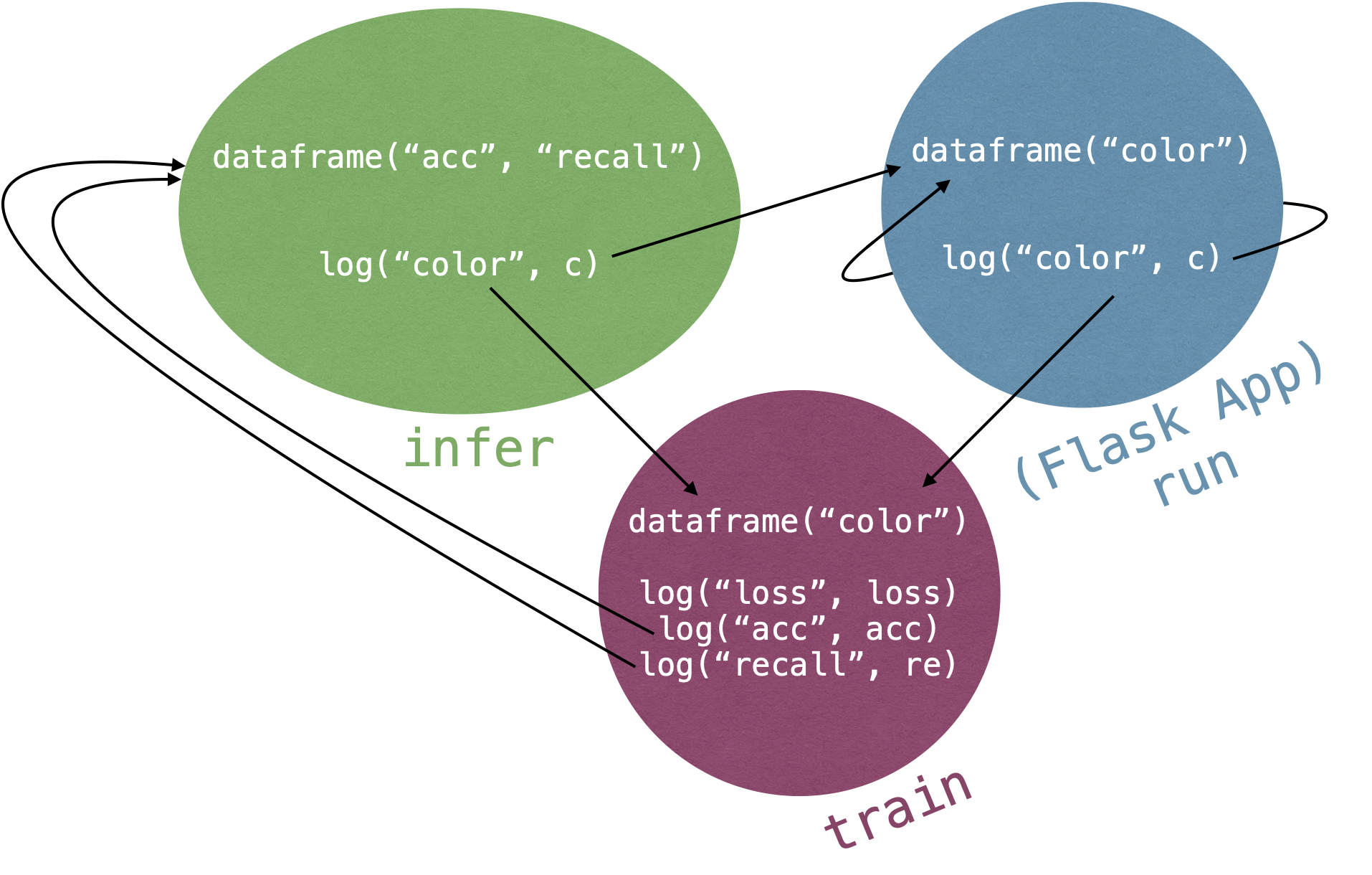}      
    \end{minipage}%
    \hspace*{-0.3cm}
    \begin{minipage}[b]{0.425\linewidth}
        \centering
        \includegraphics[width=\linewidth]{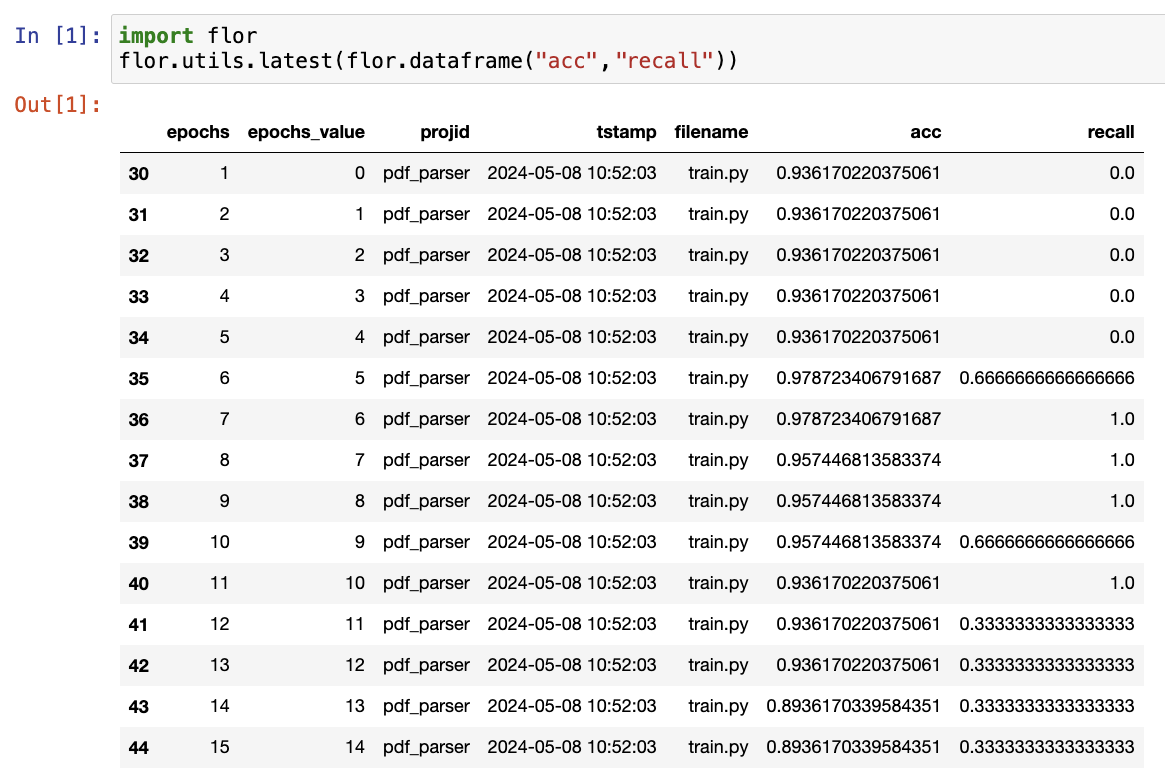}
    \end{minipage}    
    \lstset{style=custompython}

    \caption{ML Pipeline with Feedback: Makefile, Dataflow Diagram, and Flor Dataframe.}
    \label{fig:stratifiedflow}
\end{figure*}

\subsection{Application Context}

Application Context represents core information describing \textbf{what} raw data an application processes and interprets~\cite{hellerstein2017ground}. This includes all information that could be logged, such as the values of arbitrary expressions at runtime. FlorDB can capture this information post-hoc using multiversion hindsight logging (\Cref{sec:mvhl}) and manages it through a unified API. This approach provides a system that supports flexible, NoSQL-like data writes and powerful, SQL-like data reads.

FlorDB provides a straightforward interface for logging via \lstinline|flor.log(name, value)| statements, ensuring that each log entry is accompanied by crucial structured metadata such as \texttt{projid}, \texttt{tstamp}, \texttt{filename}, and \texttt{ctx\_id}. This metadata is captured at the time of import and embedded within every log entry, enabling unambiguous identification of the log's origin and context. The \texttt{ctx\_id} is generated during the initialization of \lstinline|flor.loop| and indicates the specific loop context a log entry belongs to, providing visibility into nested operations and possible cross-iteration dependencies (see \texttt{logs} and \texttt{loops} in \Cref{fig:datamodel}).

\begin{figure}[t!]
    \centering
    \begin{lstlisting}[style=custompythontiny]
for doc_name in flor.loop("document", os.listdir(...)):
    N = get_num_pages(doc_name)
    for page in flor.loop("page", range(N)):
        # text_src is "OCR" or "TXT"
        text_src, page_text = read_page(doc_name, page)
        flor.log("text_src", text_src)
        flor.log("page_text", page_text)
        
        # Run some featurization
        headings, page_numbers = analyze_text(page_text)
        flor.log("headings", headings)
        flor.log("page_numbers", page_numbers)
\end{lstlisting}
    \hspace*{-0.5cm}
    \includegraphics[width=1.1\columnwidth]{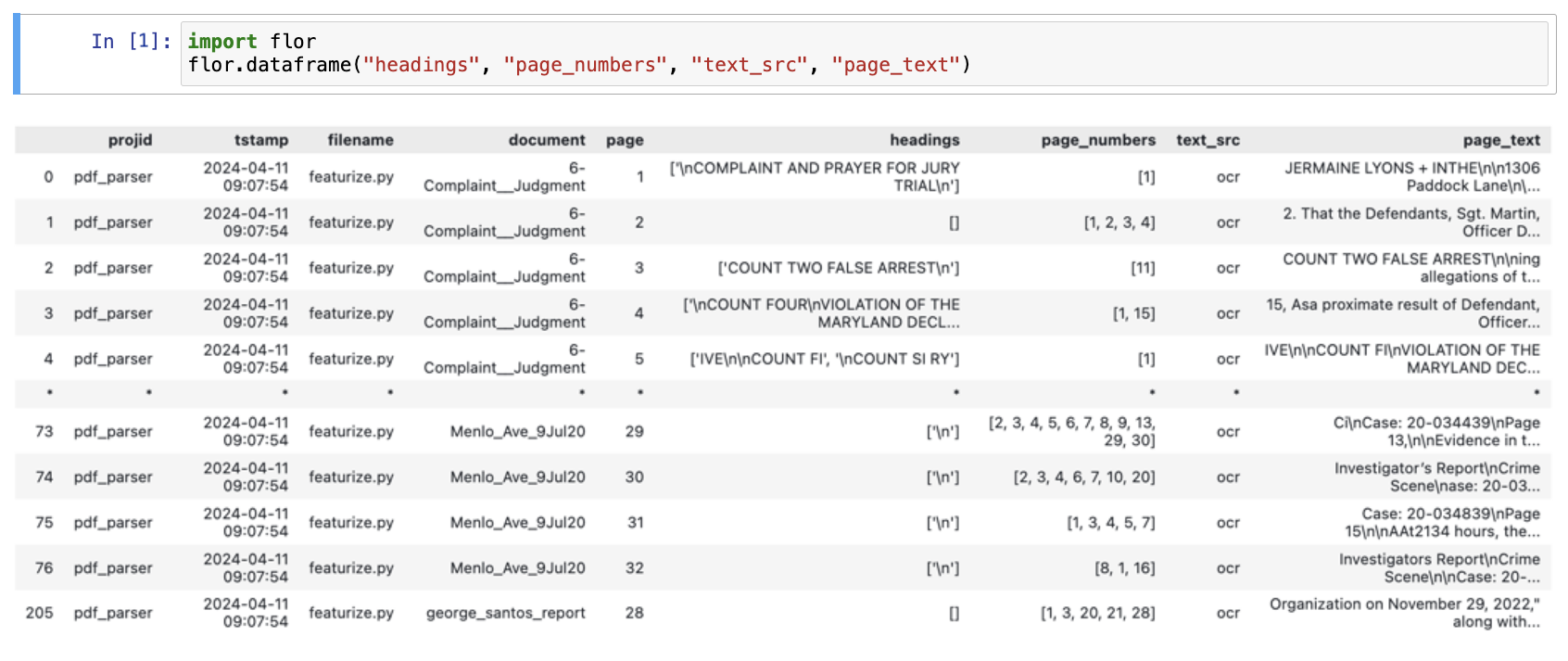}
    \caption{Data featurization with FlorDB}
    \label{fig:pdfparserFeaturize}
\end{figure}

In \Cref{fig:pdfparserFeaturize}, we give an example of how FlorDB's logging mechanism captures data features during a document analysis process. The example illustrates how multiple documents are processed, each consisting of several pages. 
This example shows how FlorDB captures a wide range of metadata without requiring a predefined schema. 
The resulting data, including the logs of headings, page numbers, text sources, and page texts, is then accessible through the \lstinline|flor.dataframe|. As shown in the bottom part of \Cref{fig:pdfparserFeaturize}, the dataframe presents this metadata in a structured layout, allowing users to query and analyze the data more efficiently~\cite{harinarayan1996implementing}.

\subsection{Behavioral \& Change Context} 
Behavioral and change context are naturally intertwined in ML pipelines: every execution both advances through the dependency graph (behavioral) and creates a new version in time (change). 
This context emerges naturally in FlorDB via logging executions with relevant file names and timestamps, rather than requiring manual documentation effort integrated into a workflow manager.

Behavioral context captures \textbf{how} data is created and used: dependency management, provenance and lineage, pipeline pathways, and dataflow. Change context tracks the version \textbf{history} of data, code, configuration parameters, checkpoints, and associated information~\cite{hellerstein2017ground}. Together, they help answer critical development questions: Where was this data defined? Where is it transformed? Who made the last change? Where should new transformations be added?

Currently, teams often manage these contexts through face-to-face interactions or communication tools like Slack or email~\cite{shankar2023we}. However, this approach creates friction by relying on colleagues' availability and memory, and doesn't scale as teams change or projects evolve. While some teams try to maintain comprehensive documentation, this conflicts with the rapid iteration typical in MLOps environments.

FlorDB manages both contexts through its logging system. When users query a \lstinline|flor.dataframe|, they receive not just data but its lineage and version history. 
To illustrate how FlorDB handles behavioral and change context, let's consider a simplified version of our document intelligence pipeline (we will cover this case in more depth in \Cref{sec:demo}).

\begin{figure*}[t]
  \centering
    \includegraphics[width=0.69\linewidth]{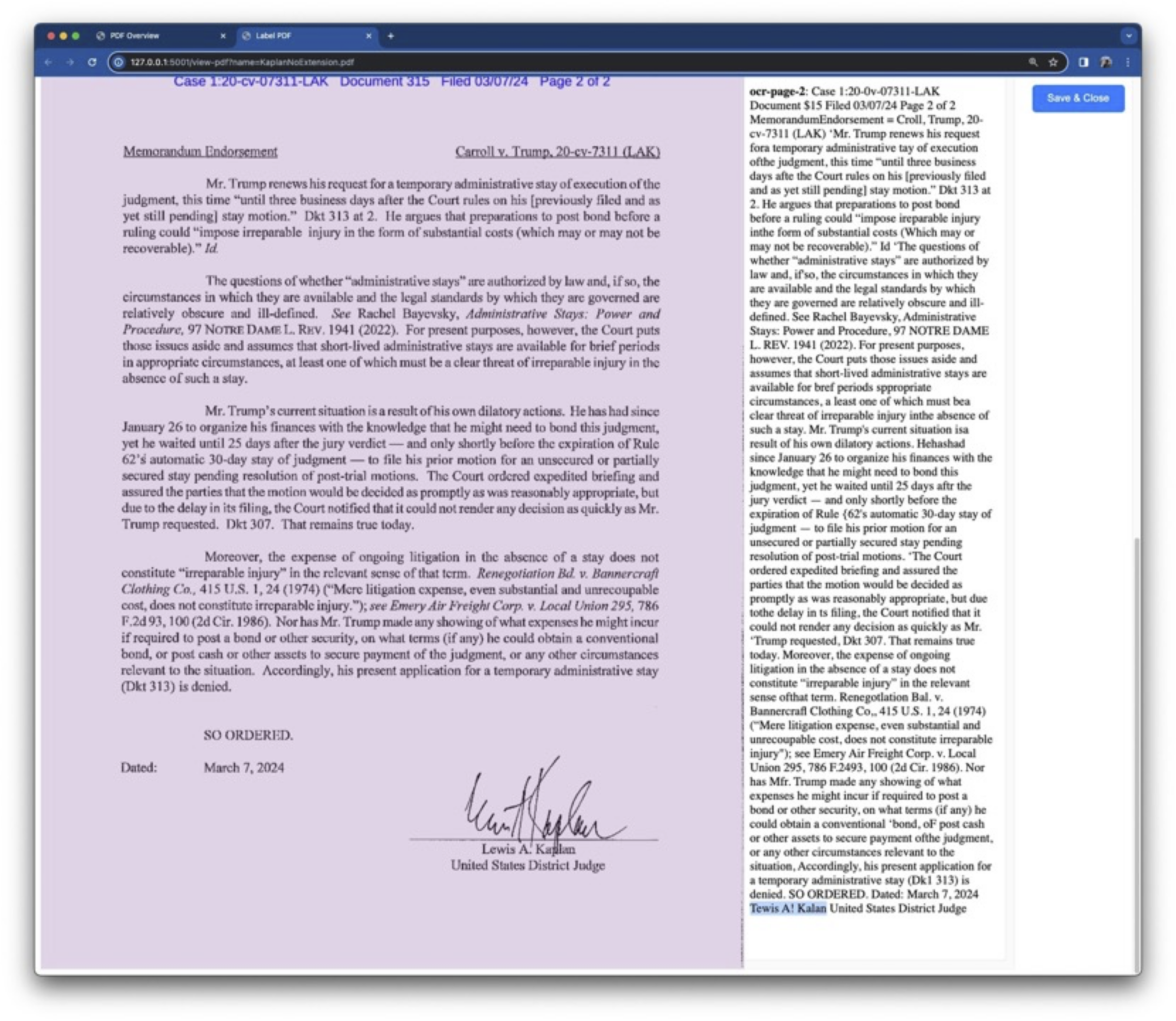}
  \hspace*{-0.2cm}
  \begin{minipage}[b]{0.3\linewidth} 
    \begin{lstlisting}[style=makefile, numbers=none]
process_pdfs: $(PDFS) pdf_demux.py
    @echo "Processing PDF files..."
    @python pdf_demux.py
    @touch process_pdfs

featurize: process_pdfs featurize.py
    @echo "Featurizing Data..."
    @python featurize.py
    @touch featurize

train: featurize hand_label train.py
    @echo "Training..."
    @python train.py

model.pth: train export_ckpt.py
    @echo "Generating model..."
    @python export_ckpt.py

infer: model.pth infer.py
    @echo "Inferencing..."
    @python infer.py
    @touch infer

hand_label: label_by_hand.py
    @echo "Labeling by hand"
    @python label_by_hand.py 
    @touch hand_label

run: featurize infer
    @echo "Starting Flask..."
    @flask run
    \end{lstlisting}
  \end{minipage}
  \caption{Screenshot of the PDF Parser (left) and its respective Makefile (right).}
\label{fig:pdfParserOverview}
\end{figure*}

Our simplified pipeline comprises three main Python scripts that interact with FlorDB:

\begin{enumerate}
    \item \texttt{train.py}: This script fine-tunes a machine learning model using the preprocessed data. It loads the dataset prepared by \texttt{prep.py}, trains the model, and saves a model checkpoint along with performance metrics like accuracy and recall. Throughout the process, it logs the training data, model parameters, and metrics, capturing the change context over time.

    \item \texttt{infer.py}: This script performs inference using the most effective model checkpoint. 
    \lstinline|flor.dataframe("acc", "recall")| is queried to retrieve the model checkpoint with the highest recall from the execution history. By accessing the logged performance metrics and version timestamps, it ensures that predictions are made using the best available model. The script then processes new data and logs the predictions for further analysis.

    \item \texttt{run.py}: This script launches a Flask web application that serves the model's predictions to end-users and collects human feedback. Users can review and correct the model's outputs, providing valuable annotations. The script logs these interactions using \lstinline|flor.log(name, value)|, capturing both the behavioral context (how data flows through the system) and the change context (how feedback updates the dataset).
\end{enumerate}

The dependencies between these scripts are specified in a Makefile, which orchestrates the execution order based on these dependencies (left pane, \Cref{fig:stratifiedflow}).
This dependency management ensures that each script runs in the correct order and that each stage has the necessary data and models from the previous stages. While we've used a Makefile for simplicity, these dependencies can also be specified using other workflow management tools like Airflow or MLflow, depending on the project's complexity and requirements.

In this pipeline, the process cycles between collecting new human-reviewed data with \texttt{run.py} and updating the model with \texttt{train.py}. The \texttt{infer.py} script ensures that the inference stage always utilizes the best-performing model checkpoint.

Figure~\ref{fig:stratifiedflow} illustrates the flow of data and transformations in the pipeline, highlighting how FlorDB captures both behavioral and change contexts. By logging with FlorDB, machine learning engineers (MLEs) not only build pipelines but also maintain a comprehensive history of data transformations and model updates. This approach reduces reliance on ad-hoc communication and documentation, allowing projects to evolve without losing crucial context.



\section{PDF Parser Demo}\label{sec:demo}
Real-world AI/ML applications are backed by ML pipelines of non-trivial complexity, often encompassing both computation and human-in-the-loop feedback.
This section gives an overview of the PDF Parser Demo, a practical application of FlorDB in document intelligence. We  demonstrate how FlorDB can be effectively used to manage context and dataflow that spans multiple asynchronous tasks that can generate a variety of metadata via computation and human feedback.

The PDF Parser\footnote{\url{https://github.com/ucbepic/pdf_parser}} is a Flask-based web application designed for efficient PDF document processing, including tasks like splitting PDFs, extracting text, and preparing data for analysis using natural language processing (NLP) techniques.
Users interact with the parser through a simple web interface, as shown in \Cref{fig:pdfParserOverview} (left). This interface allows users to navigate PDF documents and select the desired processing options.

In this demo, we aim to achieve several goals that showcase the capabilities of FlorDB:

\begin{itemize}
    \item \textbf{Demonstrate Basic Functionality:}
    \begin{itemize}
        \item \emph{Version Tracking Across Code Changes:} We will show how FlorDB's multiversion hindsight logging operates seamlessly even when the code has been refactored, ensuring that all changes are reviewable.
        \item \emph{Flexible Pipeline Modification:} The demo illustrates how new stages can be inserted into an existing pipeline or new pathways can be incorporated into a directed acyclic graph (DAG) within the Flor data model, enabling dynamic workflow evolution.
        \item \emph{Incremental Replay Execution:} By using the Flor dataframe, we can drive incremental replay execution, re-running only the parts of the workflow that have been selected by the user, thus saving time and resources.
    \end{itemize}
    \item \textbf{Showcase FlorDB Across Multiple Contexts:}
    \begin{itemize}
        \item \emph{Feature Storage and Querying After Execution:} We demonstrate how FlorDB can store and allow queries on data features post-execution without prior setup.
        \item \emph{Model Registry Functionality Post-Execution:}
        \begin{itemize}
            \item \emph{Comparative Metrics Across Models:} The demo compares performance metrics across different training runs after they have been executed.
            \item \emph{Post-Hoc Governance Enforcement:} We show how to apply governance policies retroactively to identify and handle issues like corrupted or malicious datasets (e.g., detecting a poisoned dataset).
        \end{itemize}
        \item \emph{Metric Registry and Visualization After Execution:} FlorDB acts as a repository for metrics post-execution, allowing for visualization of results---similar to generating and augmenting TensorBoard plots---even if this wasn't configured beforehand.
    \end{itemize}
\end{itemize}

Beyond highlighting the practical value of a context-rich approach, we hope this demo will serve as a reference implementation for those looking to get started with FlorDB.

\subsection{PDF Extraction \& Text Featurization}
Once the PDF is converted into text and image formats, ensuring there is one document per page, the process of featurization begins (\Cref{fig:pdfparserFeaturize}). This process typically involves text extraction, feature engineering, and vectorization.
This featurization process is essential for transforming raw PDF data into a structured form that is amenable to analysis and machine learning applications. The described methodology focuses on maximizing the information extracted from each page, ensuring that both textual and visual data contribute to the inferences made on the document.
\textbf{Takeaway:} When used in featurization contexts (\Cref{fig:pdfparserFeaturize}), FlorDB can provide the functionality of a \emph{feature store}.

\subsection{Inference Pipeline}
The inference pipeline automates the processing and analysis of images organized into document-specific folders. FlorDB's features enhance this pipeline in several key ways. First, FlorDB's comprehensive versioning and metadata tracking enables intelligent model selection. Through queries to FlorDB's unified metadata store (e.g., \lstinline|flor.dataframe("acc", "recall")|), the pipeline can automatically select the best-performing model checkpoint based on validation metrics tracked across all training runs. This eliminates the need for manual record-keeping or separate model registries.

Second, FlorDB's logging infrastructure streamlines the processing of document pages and images. As each image is processed, pre-processed, and analyzed, FlorDB captures the full context of transformations and model predictions. This includes tracking input parameters, preprocessing steps, and model outputs in a way that maintains clear lineage between the original documents and their derived predictions.

Third, when issues arise in production, FlorDB's hindsight logging capability allows developers to retroactively add logging statements to debug problems without re-running the entire pipeline. The model then makes predictions on the images, with FlorDB logging the final inferences alongside their complete provenance. \textbf{Takeaway:} When used in inference pipelines, FlorDB functions as a comprehensive \emph{model registry} (dataframe in \Cref{fig:stratifiedflow}), managing not only checkpoint selection but also model versioning, metadata tracking, and performance metrics. This enables robust model lifecycle management while maintaining clear provenance of how models evolve through training iterations.

\subsection{Training Pipeline}
\begin{figure}[t]
    \centering
    \hspace*{-0.5cm}
    \includegraphics[width=1.1\columnwidth]{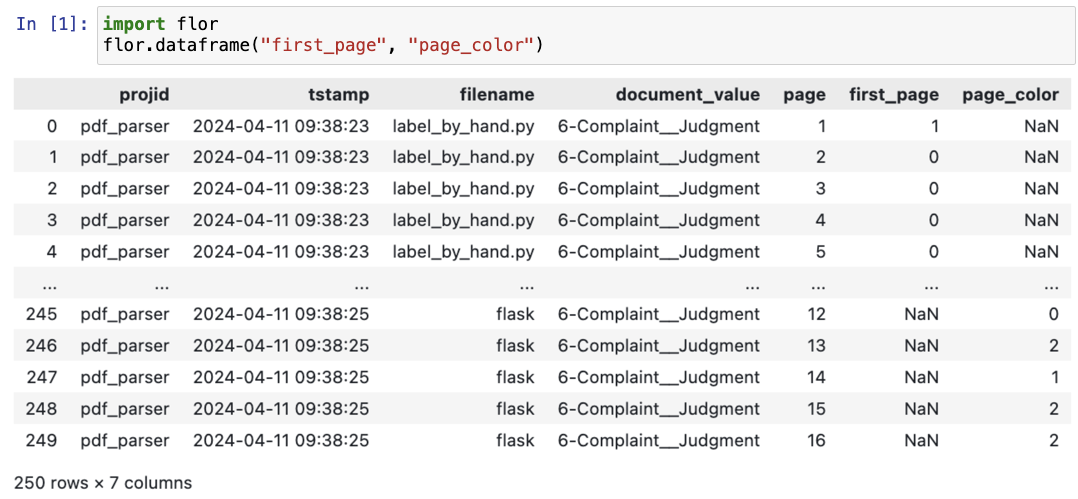}
    \begin{lstlisting}[style=custompythontiny]
labeled_data = flor.dataframe("first_page", "page_color")

hidden_size = flor.arg("hidden", default=500)
num_epochs = flor.arg("epochs", 5)
batch_size = flor.arg("batch_size", 32)
learning_rate = flor.arg("lr", 1e-3)
seed = flor.arg("seed", randint(0, 1e10))
...
with flor.checkpointing(model=net, optimizer=optimizer):
    for epoch in flor.loop("epoch", range(num_epochs)):
        for data in flor.loop("step", trainloader):
            inputs, labels = data
            optimizer.zero_grad()
            outputs = net(inputs)
            loss = criterion(outputs, labels)
            loss.backward()
            flor.log("loss", loss.item())
            optimizer.step()
        acc, recall = eval(net, testloader)
        flor.log("acc", acc)
        flor.log("recall", recall)
\end{lstlisting}
    \caption{Training on labeled data managed by FlorDB}
    \label{fig:pdfparserTrain}
\end{figure}

The training pipeline encapsulates a typical machine learning workflow, tailored for classifying images extracted from PDF pages (\Cref{fig:pdfparserTrain}).
This pipeline performs a load of training data (line 1 in \Cref{fig:pdfparserTrain}, managed by FlorDB), and data preparation. A model architecture is defined, and the model is trained over a number of epochs. 
Performance is monitored by flor logging, and context is managed and tracked using FlorDB. \textbf{Takeaway:} When used in training pipelines, FlorDB can function as a \emph{training data store} (line 1 in \Cref{fig:pdfparserTrain}), and a \emph{model repository} (lines 3-21 in \Cref{fig:pdfparserTrain}).

\subsection{Closing the Loop: Feedback via UI}
The main Flask script outlines the core functionalities of a web application designed for handling PDF documents and associated image files. It includes routes for displaying and manipulating PDFs and their converted image previews within a web interface, while also incorporating human-in-the-loop feedback for improving model performance.
The core of the application is structured around Flask routes that handle web requests and facilitate expert feedback.
Through the UI, domain experts can review model predictions and provide corrective labels, which are managed with the same metadata infrastructure as computational steps. This human feedback loop is crucial for iteratively improving model performance and maintaining data quality.

Helper functions such as \texttt{get\_colors()} fetch color data associated with the pages of a document, integrating both automated predictions and human-provided labels from the dataset (line 6 in \Cref{fig:ui-snippet}). The system maintains provenance for both machine-generated and human-provided labels, allowing developers to track the source and evolution of document annotations (top dataframe in \Cref{fig:pdfparserTrain}). The developer may choose to display labels generated by the model or labels entered manually by an expert end-user, with the metadata system capturing the origin and timestamp of each label.

\begin{figure}
    \centering
    \begin{lstlisting}[style=custompythontiny]
@app.route("/")
def home():
    return flask.render_template("index.html")

def get_colors():
    infer = flor.dataframe("first_page", "page_color")
    infer = flor.utils.latest(
        infer[infer.document_value == pdf_names[-1]])
    if infer.page_color.isna().any():
        color = infer["first_page"].astype(int).cumsum()
        infer["page_color"] = color - 1
    return infer["page_color"].to_list()

@app.route("/save_colors", methods=["POST"])
def save_colors():
    colors = request.get_json().get("colors", [])
    pdf_name = pdf_names.pop()
    with flor.iteration("document", None, pdf_name):
        for i in flor.loop("page", range(len(colors))):
            flor.log("page_color", colors[i])
    flor.commit()
    return jsonify({"message": "Colors saved"}), 200
\end{lstlisting}
\caption{Flask routes demonstrating the human-in-the-loop feedback system: The home route serves the interface, \texttt{get\_colors()} retrieves existing labels from FlorDB, and \texttt{save\_colors()} captures expert corrections while maintaining provenance through FlorDB's metadata management.}
    \label{fig:ui-snippet}
\end{figure}

\begin{itemize}
    \item The root route (``/'') displays a home page which lists all the PDF files located in a specified directory. Each PDF file is represented with a preview image, and these images are listed on the webpage using rendered HTML templates.
    \item The ``/view-pdf'' route handles requests to view a specific PDF. Depending on user interactions and the file's existence, it can display the document in different modes such as labeled text or named entity recognition (NER) views. This route also supports expert annotation interfaces where users can correct model predictions.
    \item The ``/save\_colors'' route is a POST endpoint that processes user-submitted data concerning color settings associated with a PDF's pages. This route captures this feedback data, logs it with appropriate metadata for tracking, and acknowledges the successful saving of data. The human feedback is stored with the same robust provenance tracking as computational results.
\end{itemize}

This bidirectional flow between computational processing and human expertise creates a complete feedback loop in the system. When experts provide corrections through the UI, the metadata system captures these annotations alongside the original model predictions. This human feedback can then be incorporated into subsequent model training iterations, with FlorDB maintaining clear provenance of which predictions were machine-generated versus human-corrected. This systematic approach to managing both computational and human feedback allows for continuous improvement of the model while maintaining transparency about the source and evolution of all labels in the system.

\textbf{Takeaway}: When used within human-in-the-loop interfaces, FlorDB functions as a comprehensive \emph{feedback management system} (dataframe in \Cref{fig:pdfparserTrain}), maintaining provenance for both machine-generated and human-provided labels while enabling continuous model improvement through expert corrections.

\section{Discussion}
This section discusses design features of FlorDB in the context of modern MLOps principles and best practices. We examine how FlorDB embodies the 3Vs of MLOps~\cite{shankar2023we} and discuss its design inspiration from the Ground data context service~\cite{hellerstein2017ground}.

First, we assess how FlorDB embodies and extends the 3Vs of MLOps~\cite{shankar2023we}:
\begin{itemize}
\item \textbf{Velocity}: FlorDB enhances the speed of ML development by getting metadata definition out of the critical path of agile experimentation. Rather than requiring upfront documentation that could slow iteration, FlorDB's hindsight logging capability allows teams to move fast initially while preserving the ability to retroactively capture and analyze any needed metadata on demand. This approach maintains development velocity without sacrificing the rich context needed for effective ML lifecycle management.
\item \textbf{Visibility}: The system's comprehensive logging and monitoring capabilities increase ML lifecycle transparency, facilitating debugging, optimization, and understanding of model behavior across different phases.
\item \textbf{Versioning}: FlorDB radically augments version control by providing a clean, simple interface to query metadata across versions, including technology for hindsight logging across those versions. This enables detailed tracking and allows teams to observe and understand arbitrary features of models  as they evolve over time.
\end{itemize}

Building on these MLOps principles, FlorDB's design draws inspiration from Ground's vision for data context services~\cite{hellerstein2017ground}. Like Ground, FlorDB recognizes that effective metadata management must capture not just static pre-declared descriptions but the full context of data usage across multiple dimensions. Specifically, FlorDB adapts Ground's ``ABCs of Data Context'' - Applications, Behavior, and Change - but reframes them through an MLOps lens. While Ground focused broadly on data context for analytics, FlorDB specializes these concepts for the machine learning lifecycle, treating ML models and pipelines as first-class citizens that require rich contextual tracking.

\subsection{Implications for Social Justice Research}
FlorDB was developed in the context of the Berkeley EPIC Data Lab\footnote{\url{https://epic.berkeley.edu}}, whose mission encompasses democratizing data work via no-code and low-code interfaces, informed by applications in the social justice domain including criminal defense and investigative journalism. FlorDB's design principles make it particularly well-suited for supporting social justice research efforts. By simplifying the tracking and organization of diverse datasets that social justice research often requires, FlorDB enables efficient data management even for users without extensive technical backgrounds. This accessibility is crucial for democratizing ML research capabilities across different communities and applications.

The system's ``metadata later'' approach addresses key challenges faced by resource-constrained organizations, allowing teams to focus on urgent operational demands while maintaining the ability to refine documentation post-hoc as needed. For projects involving multiple data sources, such as public health studies or legal cases, FlorDB simplifies the tracking of data movement between sources, making processes more understandable to non-specialists~\cite{epstein2019role}.

FlorDB's focus on transparency and accountability (i.e. visibility) is especially valuable for investigating algorithmic bias and fairness. The system's multiversion hindsight logging enables better tracking of decisions made during model training and development, helping researchers address concerns about the ``black box'' nature of ML models~\cite{rudin2019stop}. Moreover, its flexible metadata platform supports human-in-the-loop workflows, enabling continuous learning based on community feedback---a necessity for ethically deploying ML in dynamic social environments~\cite{holstein2019improving}.
By lowering technical barriers while enabling rigorous documentation and provenance tracking, FlorDB helps empower marginalized communities to participate more fully in ML research and development. This aligns with broader goals of democratizing access to ML technologies and ensuring their benefits extend equitably across society.


\section{Related Work}

Managing the lifecycle of ML models and associated data is crucial for efficient, reproducible workflows. FlorDB builds upon existing systems, offering comprehensive context management that integrates logging into a broader ecosystem for streamlined ML lifecycle management.

\textbf{Experiment Tracking and Version Control}: Systems like MLFlow~\cite{mlflow}, DVC~\cite{dvc}, and Weights \& Biases focus on managing experiments and ensuring reproducibility. They provide tools for tracking experiments, packaging code, and sharing models. While helpful for tracking model and data evolution, they primarily concentrate on experiment management without deeply integrating hindsight logging capabilities. As such, they do not (currently) address the challenges of context on demand that are the crux of our work.

\textbf{Model Management and Lineage}: ModelDB, Mistique, and Pachyderm emphasize version control and data lineage~\cite{modeldb, vartak2018mistique, pachyderm}. They track model lineage, capturing relationships between models, training data, and code. These systems focus on managing provenance and evolution of models and data, offering ways to query and visualize history. However, their focus is more on artifacts and less on process: for example, ModelDB does not automatically version code, and has no way of recovering missing data.

\textbf{End-to-End ML Workflows}: Systems like AWS SageMaker and Kubeflow provide comprehensive solutions for building, training, and deploying ML models~\cite{liberty2020elastic, kubeflow}. They offer tools for data labeling, model training, model hosting, and support for scalable ML workflows. Both platforms emphasize scalability and operational efficiency but primarily focus on deployment and operational aspects. FlorDB can use either system as a drop-in replacement to Make, the default build system. Other systems in this space include Helix~\cite{xin2018helix} and Motion~\cite{shankar2024building}.

\textbf{Visualization and Monitoring}: TensorBoard~\cite{tensorboard}, a visualization toolkit for TensorFlow, allows users to track and visualize metrics, graphs, and other aspects of ML experiments. FlorDB can be used with TensorBoard to visualize training metrics.

\textbf{Data Catalogs and Metadata Management}: Enterprise data catalogs like Collibra and Alation provide comprehensive solutions for organizing and discovering data assets across organizations. Collibra offers advanced governance features and automated workflows for policy management, while emphasizing data quality and observability capabilities~\cite{Collibra}. Alation focuses on intelligent data discovery through machine learning-powered search and active metadata management, helping organizations understand data lineage and trustworthiness~\cite{Alation}. In the open-source domain, DataHub and Amundsen represent modern approaches to metadata management. DataHub provides a platform spanning data discovery, observability, and governance~\cite{DataHub}, while Amundsen employs a modular architecture with dedicated services for metadata, search, and frontend interactions~\cite{Amundsen}. These systems primarily focus on metadata discovery and governance, 
helping organizations create and maintain inventories of data assets across their digital landscape. While these platforms excel at data discovery and governance, they lack FlorDB's deep integration with ML workflows and its ability to capture and recover fine-grained execution context.


\section{Conclusion}
FlorDB provides a new, general-purpose methodology and Python-based system for managing the complex metadata---the context---associated with the machine learning lifecycle. By resolving the longstanding friction between discipline and agility in metadata management, FlorDB offers AI/ML groups a flexible yet powerful solution aligning with modern MLOps agile development practices.

The key contributions of FlorDB include features that enhance the developer experience and streamline machine learning workflows. First, incremental context maintenance allows developers to gradually build out the metadata and project structure within their existing workflows. Second, the extensions presented to support pipelines, dataflow, and feedback are backward compatible with multiversion hindsight logging, allowing for a ``metadata later'' approach that enables the addition and refinement of metadata post-hoc. This capability supports rapid iteration and adaptation without compromising the long-term maintainability of projects.
As demonstrated through the PDF Parser demo, FlorDB can take on multiple roles within the ML pipeline, acting as a feature store, model registry, training data store, and experiment record. This versatility eliminates the need for multiple specialized tools while maintaining robust metadata management throughout the project lifecycle.

The development of FlorDB was informed by an interview study of ML Engineers to understand how they operationalize the machine learning lifecycle~\cite{shankar2023we}, ensuring our design decisions were grounded in real-world practices and needs. While this evidence-based design approach helped shape FlorDB's features and interface, a rigorous validation of its usability and other human factors through controlled user studies remains as important future work.

By providing a systematic way to balance agility and rigor, FlorDB creates new possibilities for scaling ML operations, improving collaboration between teams, and ensuring long-term maintainability of ML systems. As the field continues to evolve, FlorDB's flexible architecture positions it to adapt to emerging needs while maintaining its core promise: enabling teams to move fast without sacrificing reproducibility and rigor.

\bibliographystyle{ACM-Reference-Format}
\bibliography{sigproc}

\end{document}

%% file: commands.tex
\definecolor{brewerpurple}{HTML}{AF4EA3}

\definecolor{brewerlightblue}{HTML}{a6cee3}
\definecolor{brewerblue}{HTML}{1f78b4}
\definecolor{brewergreen}{HTML}{b2df8a}

\definecolor{linegray}{gray}{0.5} 

\lstdefinestyle{custompython}{
    language=Python,
    basicstyle=\ttfamily\small,
    keywordstyle=\color{brewerblue},
    morekeywords={with},
    commentstyle=\color{gray},
    stringstyle=\color{brewergreen},
    numbers=left,
    numberstyle=\tiny\color{linegray},
    numbersep=8pt,
    emph={flor,torch},emphstyle=\bfseries\color{brewerpurple},
}

\lstdefinestyle{custompythontiny}{
    language=Python,
    basicstyle=\ttfamily\footnotesize,
    keywordstyle=\color{brewerblue},
    morekeywords={with},
    commentstyle=\color{gray},
    stringstyle=\color{brewergreen},
    emph={flor,torch},emphstyle=\bfseries\color{brewerpurple},
}

\lstdefinestyle{sqlc}{
    mathescape=true,
    language=SQL,
    showspaces=false,
    showstringspaces=false,
    basicstyle=\ttfamily,
    commentstyle=\color{gray},
    morekeywords={crosstab, with}
}

\definecolor{gitgreen}{RGB}{0, 128, 0}

\makeatletter
\pgfarrowsdeclare{crow's foot}{crow's foot}
{
    \pgfarrowsleftextend{+-.5\pgflinewidth}%
    \pgfarrowsrightextend{+.5\pgflinewidth}%
}
{
    \pgfutil@tempdima=0.6pt%
    \pgfsetdash{}{+0pt}%
    \pgfsetmiterjoin%
    \pgfpathmoveto{\pgfqpoint{0pt}{-9\pgfutil@tempdima}}%
    \pgfpathlineto{\pgfqpoint{-13\pgfutil@tempdima}{0pt}}%
    \pgfpathlineto{\pgfqpoint{0pt}{9\pgfutil@tempdima}}%
    \pgfpathmoveto{\pgfqpoint{0\pgfutil@tempdima}{0\pgfutil@tempdima}}%
    \pgfpathmoveto{\pgfqpoint{-8pt}{-6pt}}%
    \pgfpathlineto{\pgfqpoint{-8pt}{-6pt}}%
    \pgfpathlineto{\pgfqpoint{-8pt}{6pt}}%
    \pgfusepathqstroke%
}

\pgfarrowsdeclare{omany}{omany}
{
    \pgfarrowsleftextend{+-.5\pgflinewidth}%
    \pgfarrowsrightextend{+.5\pgflinewidth}%
}
{
    \pgfutil@tempdima=0.6pt%
    \pgfsetdash{}{+0pt}%
    \pgfsetmiterjoin%
    \pgfpathmoveto{\pgfqpoint{0pt}{-9\pgfutil@tempdima}}%
    \pgfpathlineto{\pgfqpoint{-13\pgfutil@tempdima}{0pt}}%
    \pgfpathlineto{\pgfqpoint{0pt}{9\pgfutil@tempdima}}%
    \pgfpathmoveto{\pgfqpoint{0\pgfutil@tempdima}{0\pgfutil@tempdima}}%
    \pgfpathmoveto{\pgfqpoint{0\pgfutil@tempdima}{0\pgfutil@tempdima}}%
    \pgfpathmoveto{\pgfqpoint{-6pt}{-6pt}}%
    \pgfpathcircle{\pgfpoint{-11.5pt}{0}} {3.5pt}
    \pgfusepathqstroke%
}

\pgfarrowsdeclare{one}{one}
{
    \pgfarrowsleftextend{+-.5\pgflinewidth}%
    \pgfarrowsrightextend{+.5\pgflinewidth}%
}
{
    \pgfutil@tempdima=0.6pt%
    \pgfsetdash{}{+0pt}%
    \pgfsetmiterjoin%
    \pgfpathmoveto{\pgfqpoint{0\pgfutil@tempdima}{0\pgfutil@tempdima}}%
    \pgfpathmoveto{\pgfqpoint{-6pt}{-6pt}}%
    \pgfpathlineto{\pgfqpoint{-6pt}{-6pt}}%
    \pgfpathlineto{\pgfqpoint{-6pt}{6pt}}%
    \pgfpathmoveto{\pgfqpoint{0\pgfutil@tempdima}{0\pgfutil@tempdima}}%
    \pgfpathmoveto{\pgfqpoint{-8pt}{-6pt}}%
    \pgfpathlineto{\pgfqpoint{-8pt}{-6pt}}%
    \pgfpathlineto{\pgfqpoint{-8pt}{6pt}}%
    \pgfusepathqstroke%
}

\pgfarrowsdeclare{zeroone}{zeroone}
{
    \pgfarrowsleftextend{+-.5\pgflinewidth}%
    \pgfarrowsrightextend{+.5\pgflinewidth}%
}
{
    \pgfutil@tempdima=0.6pt%
    \pgfsetdash{}{+0pt}%
    \pgfsetmiterjoin%
    \pgfpathcircle{\pgfpoint{-11.5pt}{0}} {3.5pt}
    \pgfpathmoveto{\pgfqpoint{-6pt}{-6pt}}%
    \pgfpathlineto{\pgfqpoint{-6pt}{6pt}}%
    \pgfusepathqstroke%
}

\makeatother

\tikzset{%
    zig zag to/.style={to path={(\tikztostart) -| ($(\tikztostart)!#1!(\tikztotarget)$) |- (\tikztotarget)}},
    zig zag to/.default=0.5,
    one to zeroone/.style={
        one-zeroone, zig zag to,
    },
    one to one/.style={one-one, zig zag to},
    one to many/.style={one-crow's foot, zig zag to},
    one to omany/.style={one-omany, zig zag to},
    many to one/.style={crow's foot-one, zig zag to},
    many to many/.style={crow's foot-crow's foot, zig zag to},
    blacktable/.style={rectangle split, rectangle split parts=2, draw, align=center},
    redtable/.style={blacktable, fill=gray!20},
    font=\sffamily\small
}

\lstdefinestyle{makefile}{
    language=make,                
    basicstyle=\ttfamily\footnotesize,         
    keywordstyle=\color{blue},   
    commentstyle=\color{magenta}, 
    stringstyle=\color{magenta},  
    showstringspaces=false,      
    tabsize=4 
}